# Benchmark Study for Quantum Statevector Simulators for Scientific Computing

Timothy Myung Joon Yeo
*Software Technology Innovation Center (STIC)*
*SLB*
Sunnyvale, USA
tyeo@slb.com

Rodney Lessard
*Software Technology Innovation Center (STIC)*
*SLB*
Sunnyvale, USA
rlessard@slb.com

***Abstract*— Quantum circuit simulators are essential for quantum algorithm development, yet systematic runtime benchmarks comparing major frameworks across CPU and GPU backends are limited. We benchmark three open-source frameworks – Cirq (qsim), CUDA-Q, and Qiskit (Aer) – on statevector simulation of the 2D quantum acoustic wave equation, evolved via Trotterized Hamiltonian dynamics, for problem sizes from 7 to 29 qubits on a single Azure virtual machine. The evaluation spans four categories: CPU runtime versus qubit count, strong scalability on multi-core CPUs at 27 qubits, GPU runtime versus qubit count, and GPU-over-CPU speedup. Relative rankings shift from one category to the next. On CPU, Cirq is fastest up to 23 qubits (2.4× faster than Qiskit at 21 qubits), but Qiskit overtakes it at larger sizes and higher thread counts, sustaining 80% parallel efficiency at 16 threads. On GPU, CUDA-Q leads between 13 and 21 qubits, while Cirq becomes the fastest GPU backend at large problem sizes, achieving 1.5× and 2.3× speedups over CUDA-Q and Qiskit at 29 qubits. GPU-over-CPU speedup ratios should be interpreted alongside absolute GPU runtimes; CUDA-Q's 1500× at 21 qubits is largely driven by the limited scaling of its CPU backend (qpp), and the absolute GPU runtimes show that the three frameworks' GPU backends differ by less than these ratios alone suggest. These results indicate that no single framework dominates across configurations, and that backend selection for statevector simulation should be guided by both target hardware and problem size.**

## I. INTRODUCTION

Current quantum computers are Noisy Intermediate-Scale Quantum (NISQ) devices [1], characterized by a limited number of qubits and the absence of full error correction. While progressing toward Fault-Tolerant Quantum Computers (FTQC) continues, the vast majority of quantum algorithm development, testing, and debugging is conducted on classical circuit simulators, owing to the limited availability and high cost of quantum processing units (QPUs). Fast simulators accelerate the research cycle by enabling rapid prototyping, iteration, and validation for quantum circuits prior to hardware deployment.

Two principal classes of quantum circuit simulators execute on classical hardware. Statevector simulators model quantum computation exactly, faithfully capturing large entanglement and deep circuits. However, the memory requirement grows exponentially with the number of qubits; the largest statevector simulation reported to date reached 50 qubits on a supercomputer [2]. Tensor network-based circuit simulators, by contrast, can represent circuits with thousands of qubits when entanglement remains small to moderate, though only approximately. For example, NVIDIA researchers utilized a tensor network simulator to solve Max-Cut instances of up to 5,000 qubits on 1,688 GPUs, achieving 93% accuracy [3]. Nevertheless, as entanglement and circuit depth grow, the computational cost of tensor network contraction also scales exponentially [4].

Several open-source quantum frameworks, notably IBM's Qiskit [5], Google's Cirq [6], and NVIDIA's CUDA-Q [7], orchestrate the software and hardware stack required to develop and execute quantum algorithms. Each framework provides one or more circuit simulators targeting CPUs, GPUs or QPUs, together with libraries for circuit construction, transpilation, and analysis. Because algorithm development and validation are conducted primarily through these frameworks, simulator performance directly affects researcher productivity and iteration speed.

Prior benchmark studies have examined quantum frameworks from various angles. Nation et al. [8] introduced Benchpress, a benchmark suite of over 1,000 test cases covering seven quantum software packages, including Braket, Cirq, pytket, and Qiskit; however, their evaluation focused on circuit creation, manipulation and compilation rather than simulation runtime. Schieffer et al. [9] compared tensor network and statevector simulators within CUDA-Q on several benchmark problems, but did not extend the comparison to other frameworks. Leonteva et al. [10] benchmarked seven Matrix Product State (MPS) based and non-MPS based simulators on 13 circuit classes at utility scale (100+ qubits) on CPU backend only. To our knowledge, no existing study benchmarks statevector simulation runtime across multiple frameworks on both CPU and GPU backends – a gap this work addresses.

As a representative workload we adopt the two-dimensional acoustic wave equation via a Trotterized Hamiltonian evolution circuit [11]. The wave equation is central to the energy industry, underpinning seismic exploration, shallow-hazards identification, and carbon dioxide sequestration monitoring [12]. Quantum simulation of the wave equation benefits from exponential state-space compression: each additional qubit doubles the spatial resolution per dimension, so only three additional qubits are needed to double the grid resolution in all three spatial dimensions for a 3D problem. While forward simulation alone is unlikely to offer quantum advantage – reconstructing the full wavefield requires a number of measurements that scales exponentially with qubit count – inverse problems such as full-waveform inversion (FWI) and reverse-time migration (RTM) may provide a more plausible route to quantum speedup by computing objective functions directly, thereby circumventing costly full-state readout [13]. Regardless of the ultimate quantum-advantage question, the benchmark circuit has a large total gate count that makes it a demanding and physically motivated test for simulator performance.

We conduct all experiments on a single Azure virtual machine equipped with a multi-core CPU and an NVIDIA GPU, ensuring a controlled comparison environment. Our main contributions are as follows:

- We implement quantum circuits for running 2D acoustic wave simulation in Cirq, CUDA-Q, and Qiskit, enabling a direct comparison of statevector simulation kernel performance across all three frameworks.
- We benchmark CPU statevector simulators as a function of qubit count across all three frameworks.
- We evaluate strong scalability and parallel efficiency on multi-core CPUs for Cirq and Qiskit.
- We benchmark GPU statevector simulator performance and quantify GPU-over-CPU speedup for each framework.
- We identify performance crossover points, scaling bottlenecks, and framework-specific behaviors, providing actionable guidance for practitioners selecting a simulation backend.

## II. Methodology

### A. Quantum Circuit for 2D Acoustic Wave Simulation

The governing equation for the two-dimensional isotropic acoustic wave equation is given as follows:

$$\frac{\partial^2 u}{\partial t^2} = v^2 \nabla^2 u \tag{1}$$

where $u$ is the acoustic wavefield, $v$ is the speed of the acoustic wave through a medium, $t$ is time and $\nabla^2$ is the Laplacian operator. To simulate (1) on a quantum computer, Costa et al. [14] recast it as a Hamiltonian evolution problem by mapping it onto the time-dependent Schrödinger equation. Explicit quantum circuit constructions for the isotropic acoustic wave equation in Fourier space, including the formulation used in this work, are developed in [11].

Following the formulation in [11], the associated Schrödinger equation with Hamiltonian $H$ reads

$$\frac{\partial|\psi\rangle}{\partial t} = -iH|\psi\rangle = -i\left(X \otimes O_x + Y \otimes O_y\right)|\psi\rangle \tag{2}$$

where $X$ and $Y$ are Pauli matrices that satisfy the anticommutation relation $\{X, Y\} = 0$ required for a valid Hamiltonian representation of the 2D wave equation [11] and $O_a = \sqrt{-\nabla_a^2}, a = x, y$. The normalized state $|\psi\rangle$ encodes both the wavefield and its time derivative as follows:

$$|\psi\rangle \propto \begin{pmatrix} u \\ i\left(O_x + iO_y\right)^{-1}\partial_t u/v \end{pmatrix} \tag{3}$$

We discretize the spatial domain on a $2^n \times 2^n$ grid with periodic boundary conditions. Periodic boundaries are a natural choice, under the quantum Fourier transform (QFT), since they diagonalize the finite-difference Laplacian operator [11]. Each spatial dimension is encoded in $n$ qubits, yielding a total of $2n + 1$ qubits: $n$ for the x-dimension, $n$ for the y-dimension, and one ancilla qubit that couples the wavefield to its time derivative via the Pauli matrices in the Hamiltonian. For instance, $n = 12$ qubits per dimension corresponds to a $4{,}096 \times 4{,}096$ grid and the circuit requires 25 qubits in total.

Using the second-order central finite difference approximation for the Laplacian operator and the diagonalizing property of the QFT, the solution of (2) is

$$|\psi(t)\rangle = \left(I \otimes \mathcal{F}^{\otimes 2}\right)e^{-iHt}\left(I \otimes \mathcal{F}^{\dagger\otimes 2}\right)|\psi(0)\rangle \tag{4}$$

where $\mathcal{F}^{\otimes 2}$ is the shifted quantum Fourier transform (QFT) in 2D and $|\psi(0)\rangle$ is the initial condition.

For smooth initial conditions, the small-angle approximation simplifies the time evolution operator. The resulting unitary is decomposed using the first-order Trotter product formula [15], yielding

$$\left(e^{i\pi\Delta t X_0}\prod_{q=1}^{n} e^{i\pi\Delta t 2^{n-q}(X_0\otimes Z_q)}\, e^{i\pi\Delta t Y_0}\prod_{p=n+1}^{2n} e^{i\pi\Delta t 2^{2n-p}(Y_0\otimes Z_p)}\right)^{n_t} \tag{5}$$

where subscript 0 represents an ancilla qubit, subscripts q and p are the indices for the qubits encoding data in the x- and y-directions, respectively, $\Delta t = t/n_t$ is the time-step size, and $n_t$ is the number of time steps. The per-step Trotter error scales as $O(\Delta t^2)$ for the first-order formula [15].

The two-qubit gates in (5) are exponentials of Pauli tensor products, which are not elementary gates native to quantum hardware. Each quantum framework may employ different internal decomposition strategies. Following the staircase algorithm [16], we manually decomposed these gates into elementary single-qubit rotations and CNOT gates.

For the $X \otimes Z$ term, the decomposition uses the identity $HXH = Z$ [16] to convert $X$ into $Z$ via Hadamard gates H, yielding a pair of CNOT gates surrounding an $R_Z$ rotation bracketed by Hadamard gates on the ancilla qubit:

$$\prod_{q=1}^{n} e^{i\pi\Delta t 2^{n-q}(X_0\otimes Z_q)} = \left(H_0 \otimes I^{\otimes n}\right) \left(\prod_{q=1}^{n} \mathrm{CNOT}(0,q)\left(I_0 \otimes R_{Z,q}(-\pi\Delta t 2^{n-q+1})\right)\mathrm{CNOT}(0,q)\right) \left(H_0 \otimes I^{\otimes n}\right) \tag{6}$$

where $I^{\otimes n}$ denotes the identity operator on all n spatial qubits in the respective dimension, $CNOT(0,q)$ is the CNOT gate controlled by the ancilla (qubit 0) as control and qubit q as target, and $R_{Z,q}(\theta)$ denotes an $R_Z(\theta) = e^{-i\theta/2Z}$ rotation applied to qubit q. All operators act as identity on qubits not explicitly specified.

For the $Y \otimes Z$ term, the decomposition uses the identity $R_x\left(-\frac{\pi}{2}\right) Z R_x\left(\frac{\pi}{2}\right) = Y$ to handle the Pauli Y gate via $R_x\left(\pm\frac{\pi}{2}\right)$ rotation gates on the ancilla qubit:

$$\prod_{p=n+1}^{2n} e^{i\pi\Delta t 2^{2n-p}(Y_0\otimes Z_p)} = \left(R_{X,0}(-\pi/2)\otimes I^{\otimes n}\right) \left(\prod_{p=n+1}^{2n} CNOT(0,p)\left(I_0\otimes R_{Z,p}\left(-\pi\Delta t 2^{2n-p+1}\right)\right) CNOT(0,p)\right) \left(R_{X,0}(\pi/2)\otimes I^{\otimes n}\right) \quad (7)$$

where $R_{X,0}(\theta) = e^{-i\theta/2X}$ denotes an $R_X$ rotation applied to the ancilla qubit, and $R_{Z,p}(\theta)$ is applied to qubit p. For the benchmark study, the simulation is run until $t = 1.0$ with $\Delta t = 0.01$, yielding $n_t = 100$ Trotter layers.

Fig. 1 shows a quantum circuit for 2D isotropic acoustic wave simulation with a total of 7 qubits ($n = 3$ for each spatial dimension). The circuit consists of three parts: (i) a state preparation step, which encodes the initial condition onto the amplitudes of the quantum states, (ii) a time update step, which applies the shifted QFT and its adjoint together with the Trotter-decomposed time-stepping operators from (5)–(7); and (iii) measurement or statevector retrieval. The first qubit (0-th) is the ancilla qubit that couples the wavefield and its time derivative through the Pauli matrices in the Hamiltonian in (2). The next $n = 3$ qubits (first through n-th) encode data in the x-direction and the remaining n qubits ($n + 1$-th to $2n$-th) encode data in the y-direction. Therefore, the circuit requires a total of $(2 \times n + 1)$ qubits to simulate $2^n \times 2^n$ grid points.

The initial wavefield is calculated from the 2D Ricker wavelet [17] and the initial time derivative is set to zero:

$$u(x,y,0) = \frac{1}{\pi\sigma^4}\left(1 - \frac{1}{2\sigma^2}\left((x-\mu_x)^2 + \left(y-\mu_y\right)^2\right)\right) \times \exp\left(-\frac{(x-\mu_x)^2 + \left(y-\mu_y\right)^2}{2\sigma^2}\right).$$

$$\partial u_t(x,y,0) = 0 \quad (8)$$

with $\mu_x = \mu_y = 0.5$ and $\sigma = 0.1$ and $x, y \in [0,1]$. The initial condition is normalized and then encoded into the amplitudes of the statevector. To encode the initial condition, one can use exact gate-based state preparation or parameterized quantum circuits (PQCs). The former deterministically constructs the target state but requires $(2^{n+2} - 4n - 4)$ CNOT gates and $(2^{n+2} - 5)$ one-qubit elementary rotation gates [18], while the latter can produce shallower circuits at the cost of a classical optimization loop [12].

For this study, the initial quantum state was prepared by directly setting the statevector to the desired initial condition, bypassing gate-based initialization. This capability is available only on circuit simulator backends and not on physical quantum hardware. We deliberately leveraged this feature to enable a direct evaluation of each framework's statevector simulator performance, ensuring that the benchmark results reflect pure simulation efficiency, uninfluenced by state-preparation gate operations or compilation processes.

While simulators also support measurement via sampling (using a specified number of shots), direct statevector access enables precise evaluation of the simulation engine's raw computational performance. This approach is especially suitable for benchmarking simulator backends, where the complete quantum state is internally available and can be accessed without loss of fidelity.

## B. Benchmark Configuration

We evaluated three widely used quantum computing frameworks: Cirq, CUDA-Q, and Qiskit. TABLE I. summarizes the package versions used in the benchmark. For Cirq, we utilized the qsim simulator, a high-performance statevector simulator optimized for multi-threaded CPU execution and GPU acceleration. For Qiskit, we employed the Qiskit Aer statevector simulator, which similarly supports both CPU and GPU backends. For GPU-based simulation, both Cirq and Qiskit offer cuQuantum-based simulators; however, we employed their native CUDA-based GPU simulator to compare performance. CUDA-Q leverages the Quantum++ (qpp) backend [19] for CPU-based statevector simulation and the cuStateVec backend for GPU-based statevector simulation. For Cirq and Qiskit, the official PyPI was used for CPU benchmarks, and a CUDA-enabled build from source for GPU benchmarks. For CUDA-Q, the official PyPI was used for both CPU and GPU benchmarks.

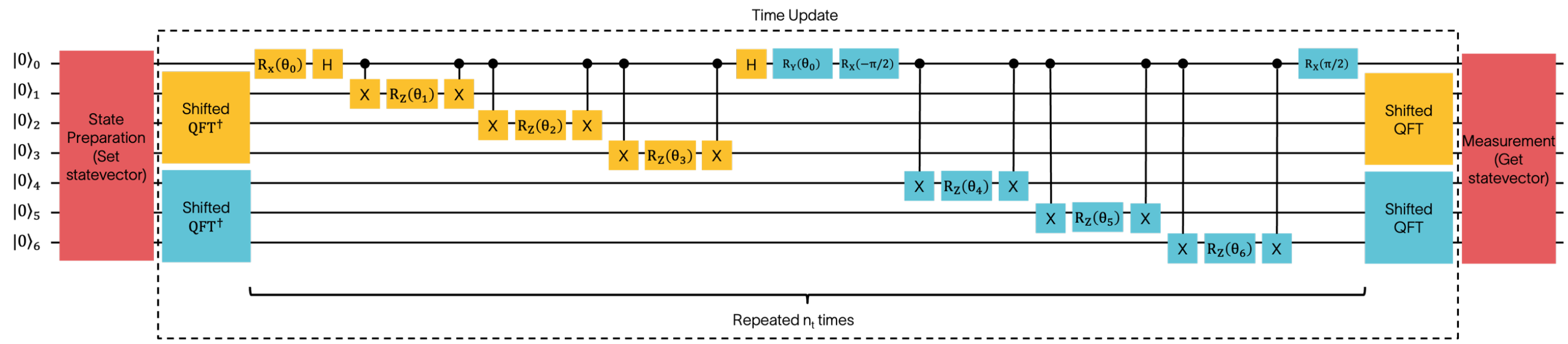

Fig. 1. Quantum circuit for the 2D acoustic wave simulation with 7 qubits. The circuit comprises three parts: state preparation, time update, and measurement. Only a single Trotter layer is shown for the time update part. Qubit 0 serves as the ancilla that couples the wavefield with its time derivative. The next $n = 3$ qubits ($q = 1,2,3$) encode the data in the x-direction, and remaining n qubits ($p = 4,5,6$) encode the data in the y-direction. The rotation angle for $R_X$ and $R_Y$ gates is $\theta_0 = -2\pi\Delta t$ and rotation angles for $R_Z$ gates are $\theta_q = -\pi\Delta t 2^{n-q+1}$, for $q = 1,2,3$ and $\theta_p = -\pi\Delta t 2^{2n-p+1}$ for $p = 4,5,6$.

Circuit execution time was measured comprehensively, encompassing circuit generation, transpilation (where applicable), and actual simulation runtime. For cases where transpilation was necessary, as with Qiskit, the optimization level was explicitly set to zero. This ensured that the compiled circuit preserved the original structure used in the other frameworks, preserving consistency across all benchmarks.

All results presented in this paper were obtained using an Azure Standard_NC24ads_A100_v4 virtual machine, equipped with an AMD EPYC 7V13 CPU (24 cores) with 220 GB of memory and an NVIDIA A100 GPU with 80 GB of memory. All CPU benchmarks were executed on a single-NUMA node system with 24 physical cores. To minimize memory bandwidth contention and ensure reproducible results, benchmarks were limited to 16 threads. Each benchmark configuration was executed 11 times. To mitigate cold-start effects, the first run was discarded as a warm-up iteration. Performance statistics, including mean and standard deviation, were computed from the remaining 10 runs. All simulations were run with single precision (complex64).

TABLE I. PACKAGE VERSIONS OF VARIOUS QUANTUM FRAMEWORKS

| Quantum SDK | Package Versions |
|---|---|
| Cirq | Cirq==1.6.1<br>Qsimcirq==0.22.0 |
| CUDA-Q | cudaq==0.13.0 |
| Qiskit | qiskit==2.3.0<br>qiskit-aer==0.17.2 |

### C. Endianness Conventions in Quantum Frameworks

Various quantum frameworks use different default endianness conventions. There are two parts of the system where endianness can vary: qubit ordering and statevector indexing, as shown in TABLE II.

For qubit ordering, Cirq and CUDA-Q use the big-endian convention, writing the first qubit in the leftmost position of the ket, denoted as $|q_0q_1q_2\rangle$. Conversely, Qiskit employs the little-endian convention, writing the first qubit in the rightmost position, represented as $|q_2q_1q_0\rangle$.

For statevector indexing, both Cirq and Qiskit use the big-endian format, in which the leftmost qubit is the most significant bit. For example, in a three-qubit system, the statevector ordering is: $|000\rangle, |001\rangle, |010\rangle, |011\rangle, |100\rangle, |101\rangle, |110\rangle, |111\rangle$. In contrast, CUDA-Q employs the little-endian indexing, in which the first (leftmost) qubit is the least significant bit, resulting in the ordering: $|000\rangle, |100\rangle, |010\rangle, |110\rangle, |001\rangle, |101\rangle, |011\rangle, |111\rangle$.

To ensure a fair comparison across quantum frameworks, our benchmarking study preserves the original circuit structure as depicted in Fig. 1. However, appropriate pre- and post-processing transformations are applied to account for differences in endianness conventions among the frameworks. The original circuit was developed for CUDA-Q. For Cirq, the initial condition is reordered from little-endian to big-endian before encoding into the statevector, and the retrieved statevector is reordered back to little-endian to match the result by CUDA-Q. For Qiskit, the little-endian qubit ordering and big-endian statevector indexing effectively cancel each other relative to CUDA-Q, requiring no pre- or post-processing transformation. By implementing these adjustments, we ensure that each framework executes the same circuit and produces identical results, regardless of their respective internal representations. The circuit run time measured in our benchmarks excludes the time spent in pre- and post-processing.

TABLE II. QUBIT AND AMPLITUDE ORDERING ENDIANNESS OF VARIOUS QUANTUM FRAMEWORKS

| Quantum SDK | Qubit Ordering | Statevector Indexing |
|---|---|---|
| Cirq | Big-Endian | Big-Endian |
| CUDA-Q | Big-Endian | Little-Endian |
| Qiskit | Little-Endian | Big-Endian |

## III. BENCHMARK RESULTS

### A. Accuracy Comparison with Classical Simulator

Fig. 2 compares wavefield snapshots at time steps $t = 0.0, 0.3, 0.6, 0.9, 1.0$ for $2^{10} \times 2^{10}$ grid points, computed by (a) a classical finite-difference simulator using the pseudo-spectral method [20] and (b) the quantum circuit simulator. The classical simulator uses a time step of $\Delta t = 0.0001$, whereas the quantum circuit simulator uses $\Delta t = 0.01$. Despite this two-order-of-magnitude difference in time step size, the quantum simulation results closely match the classical reference solution, validating the accuracy of the quantum approach. For the classical simulator with the explicit time discretization, the time step is constrained by the Courant-Friedrichs-Lewy (CFL) stability condition [21]. For the given case using the pseudo-spectral method with the second-order explicit time discretization, the CFL condition is given by $\frac{v\Delta t}{\Delta x} \leq \frac{2}{\pi\sqrt{2}}$, which requires $\Delta t \leq 4.4 \times 10^{-4}$. The quantum circuit simulation, however, is not subject to the CFL restriction and can therefore use time steps that are 22 times larger while maintaining comparable accuracy. At $t = 1.0$, the maximum element-wise difference between any pair of frameworks is on the order of $10^{-8}$, corresponding to a relative difference of approximately $10^{-5}$ with respect to the solution magnitude. This is consistent with accumulated single-precision rounding errors over 100 Trotter steps, confirming that the three frameworks produce effectively identical results.

### B. Benchmark Results for the CPU-based Statevector Simulator

Fig. 3 presents the results of the CPU benchmark, with the number of qubits varied from 7 (corresponding to an 8×8 grid) up to 29 (16,384×16,384 grid points). All simulations were executed using 16 CPU threads. For Cirq and Qiskit, the thread count was configured via framework-specific options, which each backend then adjusts internally based on problem size; for CUDA-Q, multi-threading was controlled by setting the environment variable OMP_NUM_THREADS to 16 for all problem sizes.

Cirq is the fastest framework up to 23 qubits, with the largest advantage in the medium-scale regime (19–23 qubits). At 21 qubits, Cirq completes the circuit in 0.451 s compared to 1.07 s for Qiskit (2.4× faster) and at 23 qubits the gap widens to 2.8× (Cirq: 1.16 s, Qiskit: 3.24 s). Cirq's runtime is also noticeably noisier than the other backends across small- and medium-scale regimes: between 7 and 19 qubits, its standard deviation across runs stays between 0.0220 and 0.0252 s, roughly 4.6× larger than Qiskit's maximum of 0.00553 s in the same range.

Beyond 23 qubits, Qiskit overtakes Cirq. The crossover occurs between 23 and 25 qubits: Qiskit is 1.2× faster at 25 qubits (12.8 s vs. 14.9 s), and its advantage widens with problem size: at 27 qubits, Qiskit finishes in 67.4 s versus 83.1 s for Cirq, and at 29 qubits in 294 s versus 367 s.

CUDA-Q's CPU backend (qpp) shows pronounced performance degradation beyond 9 qubits. It is competitive with Cirq and Qiskit in the small-scale regime (7–9 qubits) but slows sharply with size, reaching 0.753 s at 13 qubits (5.2× slower than Cirq) and 74.2 s at 19 qubits. At 21 qubits, CUDA-Q requires 370 s – 820× slower than Cirq and 350× slower than Qiskit. Larger problem sizes were not tested due to prohibitive runtimes.

### C. *Strong Scalability Test for CPU-based Statevector Simulator*

Fig. 4 and TABLE III. present the strong scalability results for Cirq and Qiskit with a fixed problem size of 27 qubits. CUDA-Q is excluded because of its poor performance at this scale as shown in Fig. 3.

On a single thread, Cirq is the faster framework, completing the circuit in 251 s versus 865 s for Qiskit (3.4× faster). The two frameworks scale very differently, however, as thread count grows from 1 to 16. At 2 threads, both retain high parallel efficiency (Cirq 94%, Qiskit 97%), but their behavior diverges beyond that point. Cirq's efficiency drops sharply – to 69% at 4 threads, 38% at 8 threads, and 19% at 16 threads – yielding only a 3.0× speedup at 16 threads. The runtime essentially saturates beyond 8 threads (83.5 s vs. 82.9 s), indicating that qsim's parallel kernel does not effectively use additional threads at this problem size. Qiskit, by contrast, retains 80% efficiency at 16 threads (13× speedup), and its kernel time of 67.2 s is 1.2× faster than Cirq's 82.9 s – a complete reversal of the single-thread ranking.

This crossover is consistent with the CPU benchmark in Fig. 3, where Qiskit overtakes Cirq above 23 qubits: Qiskit's higher single-threaded overhead is progressively offset by its more effective multi-threading as the workload grows, eventually surpassing Cirq's faster but less scalable kernel.

TABLE III. RUNTIME IN SECONDS FOR CPU STRONG SCALABILITY TEST

| Number of Threads | 1 | 2 | 4 | 8 | 16 |
|---|---|---|---|---|---|
| Cirq | 251 | 134 | 91.4 | 83.5 | 82.9 |
| Qiskit | 865 | 445 | 232 | 124 | 67.2 |

### D. *Benchmark Results for the GPU-based Statevector Simulator*

Fig. 5 shows GPU benchmark results for the 2D acoustic wave simulation, with qubit counts ranging from 7 to 29. All three frameworks were run in single precision (complex64). Cirq uses qsim's CUDA backend (use_gpu=True, gpu_mode=0); Qiskit uses Aer's GPU statevector method; and CUDA-Q uses the NVIDIA cuStateVec backend (target=nvidia).

The relative ordering of the three backends changes twice across the range. Cirq is the fastest from 7 to 11 qubits, after which CUDA-Q takes the lead between 13 and 21 qubits. At 19 qubits, for example, CUDA-Q completes in 0.176 s versus 0.229 s for Cirq (1.3× faster). A second crossover occurs between 21 and 23 qubits, where Cirq overtakes CUDA-Q again and retains the lead through 29 qubits. The Cirq advantage grows with problem size, reaching 1.5× at 29 qubits (Cirq: 19.1 s, CUDA-Q: 28.2 s). As in the CPU benchmark (Fig. 3), Cirq's runtime is noticeably noisier than the other backends from 7 to 23 qubits, with run-to-run standard deviations of roughly 0.0177–0.0244 s, which is roughly an order of magnitude larger than CUDA-Q (0.00128–0.00589 s) and Qiskit (0.000928–0.00838 s) over the same range.

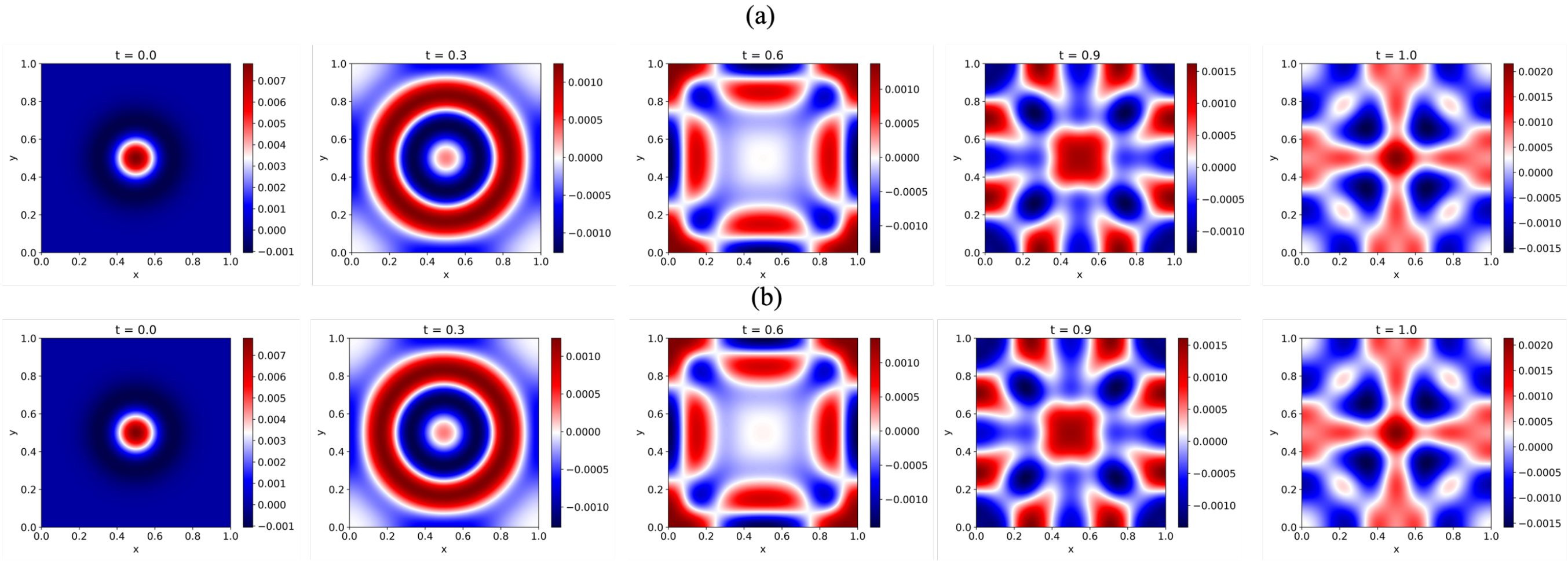

Fig. 2. Wavefield snapshots at time steps $t =$ 0.0, 0.3, 0.6, 0.9, 1.0 from (a) a classical simulator and (b) the quantum circuit simulator.

Qiskit's GPU backend is the slowest at every qubit count tested, consistently trailing both CUDA-Q and Cirq. At 29 qubits, Qiskit (44.8 s) is 1.6× slower than CUDA-Q and 2.3× slower than Cirq.

### E. *GPU Speedup Over CPU*

Fig. 6 presents the GPU speedup of each framework, defined as the ratio of CPU runtime to GPU runtime at the same qubit count. Values less than 1.0 indicate the GPU implementation is slower than CPU.

Both Cirq and Qiskit show limited GPU benefit at small-to-medium problem sizes (7–17 qubits): Cirq's speedup stays in the 1.0–1.1× range, and Qiskit's varies between 0.95× and 1.5×, with the GPU occasionally slower than the CPU. From 19 qubits onward, GPU acceleration becomes consistently beneficial for both frameworks.

Cirq's GPU speedup grows with problem size: 1.2× at 19 qubits, 2.3× at 23 qubits, 11× at 25 qubits, and 19× at 29 qubits. This steep acceleration reflects the widening gap between the exponentially growing memory bandwidth and compute demands of statevector simulation and the CPU's limited throughput, while Cirq's GPU backend scales more efficiently.

CUDA-Q demonstrates significant GPU speedup at substantially smaller problem sizes: 5.8× at 13 qubits, 100× at 17 qubits, and 1500× at 21 qubits, exceeding three orders of magnitude. CUDA-Q's CPU/GPU comparison is limited to qubit counts ≤ 21 because the CPU baseline (qpp) was discontinued beyond this point due to prohibitive runtimes (Fig. 3). These large speedup ratios therefore reflect a weak CPU baseline as much as a strong GPU backend.

Qiskit exhibits the most modest GPU acceleration, reaching 6.6× at 29 qubits. Because Qiskit's CPU backend is the fastest of the three at this scale (Fig. 3), the small ratio is unlikely to stem from a strong CPU baseline alone; it more likely indicates that Qiskit Aer's GPU statevector method has additional headroom for optimization. At 21 qubits, for instance, CUDA-Q and Cirq achieve essentially the same GPU runtime (0.254 s vs. 0.272 s), yet their reported speedups differ by nearly three orders of magnitude (1500× vs. 1.7×) because CUDA-Q's CPU baseline is far slower. Readers should therefore interpret Fig. 6 alongside the absolute GPU runtimes in Fig. 5: the speedup ratio measures the combined effect of GPU strength and CPU weakness, not GPU performance alone.

Fig. 7 presents GPU speedup relative to the best CPU backend at each qubit count, rather than each framework's own CPU baseline. The best CPU performer is Cirq for 7–23 qubits, and Qiskit for 25 qubits and beyond (Fig. 3). This metric isolates GPU benefit from the weak-CPU-baseline effect that inflates the per-framework ratios in Fig. 6.

CUDA-Q first surpasses the best CPU backend at 13 qubits (1.1×) and improves gradually to 2.2× at 23 qubits. Beyond this point the speedup rises sharply, reaching 7.0× at 25 qubits and 10× at 29 qubits.

Cirq overtakes the best CPU backend slightly later, at 17 qubits (1.1×), but its speedup climbs more steeply than CUDA-Q's: 1.7× at 21 qubits, 9.8× at 25 qubits, and 15× at 29 qubits.

Qiskit shows the latest crossover, exceeding the best CPU backend at 21 qubits (1.1×), and then increases moderately to 4.5× at 25 qubits, and 6.6× at 29 qubits. Because Qiskit is itself the best CPU backend at these sizes, this curve coincides with its per-framework ratio in Fig. 6, providing a consistency check between the two figures.

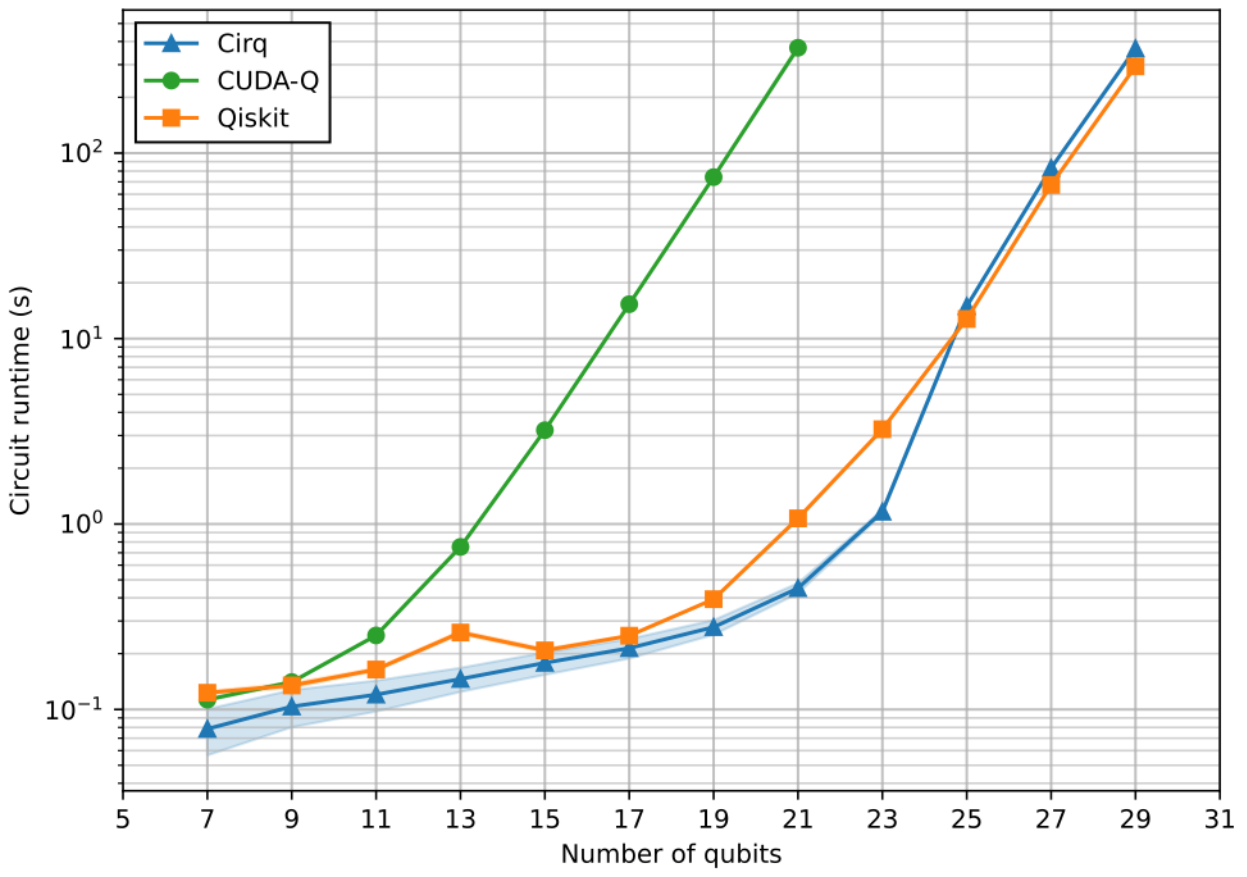

Fig. 3. Average circuit runtime as a function of qubit count for the 2D quantum acoustic wave simulation on CPU backends — Cirq (qsim), Qiskit (Aer), and CUDA-Q (qpp). All runs used 16 CPU threads on the same Azure VM. The y-axis is log-scaled. Shaded bands denote ±1 standard deviation across repeated runs. CUDA-Q data points stop at 21 qubits due to prohibitive runtime at larger problem sizes.

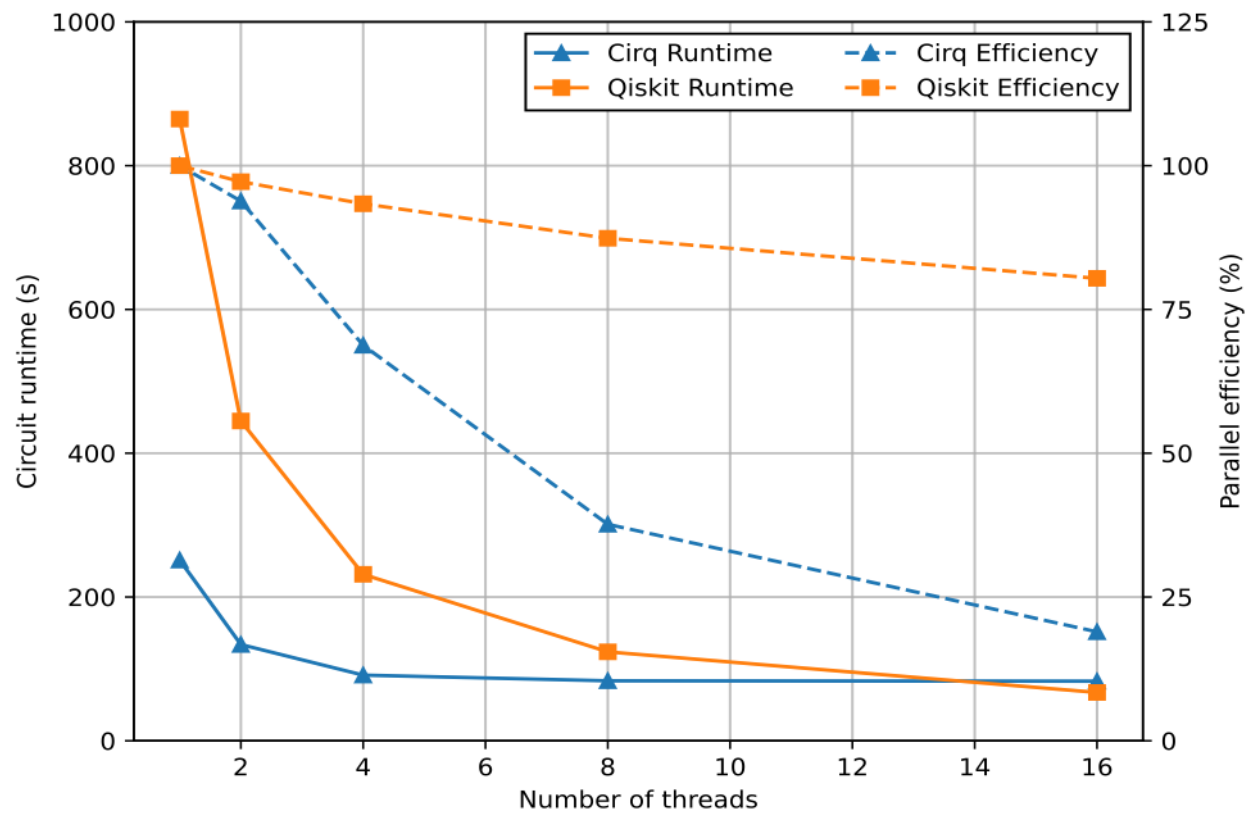

Fig. 4. Strong scalability of Cirq and Qiskit on the 2D quantum acoustic wave simulation at a fixed problem size of 27 qubits. Solid lines show the average circuit runtime (left axis); dashed lines show the parallel efficiency (right axis). The thread count is varied across 1, 2, 4, 8, and 16. The maximum coefficient of variation across all configurations is 1.02%. CUDA-Q is omitted due to its poor performance at this problem size.

The steepest rise in GPU speedup for all three frameworks occurs between 23 and 25 qubits, and it originates on the CPU side rather than from any change in GPU behavior: adding two qubits quadruples the statevector, yet every GPU backend stays at or below the ideal 4× over this step (Cirq 2.5×, CUDA-Q 3.4×, Qiskit 3.1×). The best CPU runtime instead rises about 11× (1.16 s to 12.8 s), though this figure partly reflects the baseline switching from Cirq to Qiskit at 25 qubits. Holding the framework fixed, the asymmetry is clear: Cirq's CPU kernel degrades by 12.8× over this step while Qiskit's grows only 3.9×, close to the ideal factor. The sharp speedup is therefore driven mainly by Cirq's steep, framework-specific slowdown, whose qsim kernel also ceases to benefit from additional threads at large statevectors (Fig. 4), rather than by a uniform CPU limit affecting both backends. In all cases the GPUs widen their advantage precisely where the CPU backends falter.

## IV. Summary and Future Work

### A. Summary

This study evaluated the performance of three quantum computing frameworks – Cirq, CUDA-Q, and Qiskit – for statevector simulation of 2D acoustic wave equations, across CPU, GPU, and multi-threaded configurations with qubit counts ranging from 7 to 29.

On CPU (Fig. 3), Cirq is the fastest framework up to 23 qubits, with the largest advantage in the medium-scale regime: 2.4× faster than Qiskit at 21 qubits and 2.8× faster at 23 qubits. The two frameworks cross over between 23 and 25 qubits, after which Qiskit takes the lead; at 29 qubits, Qiskit (294 s) is 1.2× faster than Cirq (367 s). CUDA-Q's CPU backend (qpp) exhibits severe performance degradation beyond 9 qubits, reaching 820× slower than Cirq at 21 qubits, and was excluded from larger CPU runs.

Strong scalability testing at 27 qubits (Fig. 4) reveals contrasting parallel characteristics between Cirq and Qiskit. Cirq begins with a 3.4× single-threaded advantage but its parallel efficiency drops sharply to 19% at 16 threads, yielding only a 3.0× speedup. Qiskit, in contrast, sustains 80% efficiency at 16 threads (13× speedup). The result is a complete reversal of the single-thread ranking: at 16 threads, Qiskit (67.2 s) is 1.2× faster than Cirq (82.9 s). This crossover is consistent with the CPU benchmark, where Qiskit also overtakes Cirq above 23 qubits.

On GPU (Fig. 5), the performance landscape changes again. CUDA-Q delivers the fastest circuit execution between 13 and 21 qubits, while Cirq leads at both small (7–11 qubits) and large (23–29 qubits) problem sizes. At 29 qubits, Cirq (19.1 s) is 1.5× faster than CUDA-Q (28.2 s) and 2.3× faster than Qiskit (44.8 s). Qiskit's GPU backend is consistently the slowest across all tested scales.

GPU-over-CPU speedup analysis (Fig. 6) highlights framework-specific tradeoffs. CUDA-Q achieves the largest speedup ratios, 1500× at 21 qubits, but this primarily reflects its weak CPU baseline rather than an exceptionally strong GPU backend: at 21 qubits, CUDA-Q's GPU runtime (0.254 s) is essentially indistinguishable from Cirq's (0.272 s), despite their speedup ratios differing by nearly three orders of magnitude. Cirq shows little GPU benefit below 19 qubits but reaches 19× speedup at 29 qubits, indicating that its CPU backend remains competitive until large statevectors expose the throughput advantage of GPU execution. Qiskit's GPU speedup reaches only 6.6× at 29 qubits, suggesting room for optimization in Aer's GPU statevector method.

Normalizing each GPU backend against the best CPU backend at every qubit count (Fig. 7) removes the weak-baseline distortion of Fig. 6 and yields a fairer cross-framework comparison. Under this metric the GPU first overtakes the best

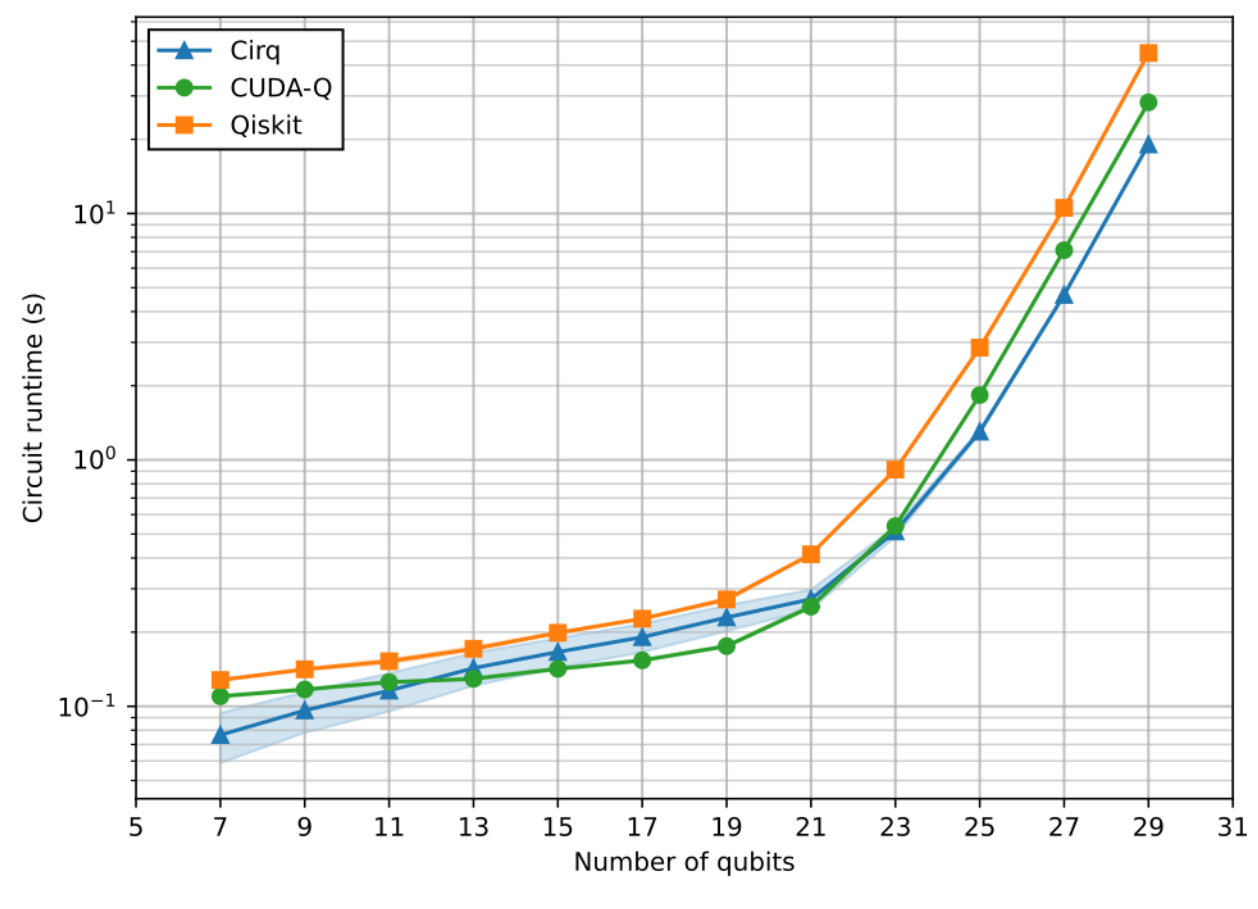

Fig. 5. Average circuit runtime as a function of qubit count for the 2D quantum acoustic wave simulation on GPU backends – Cirq (qsim), Qiskit (Aer), and CUDA-Q (cuStateVec). The y-axis is log-scaled. Shaded bands denote ±1 standard deviation across repeated runs.

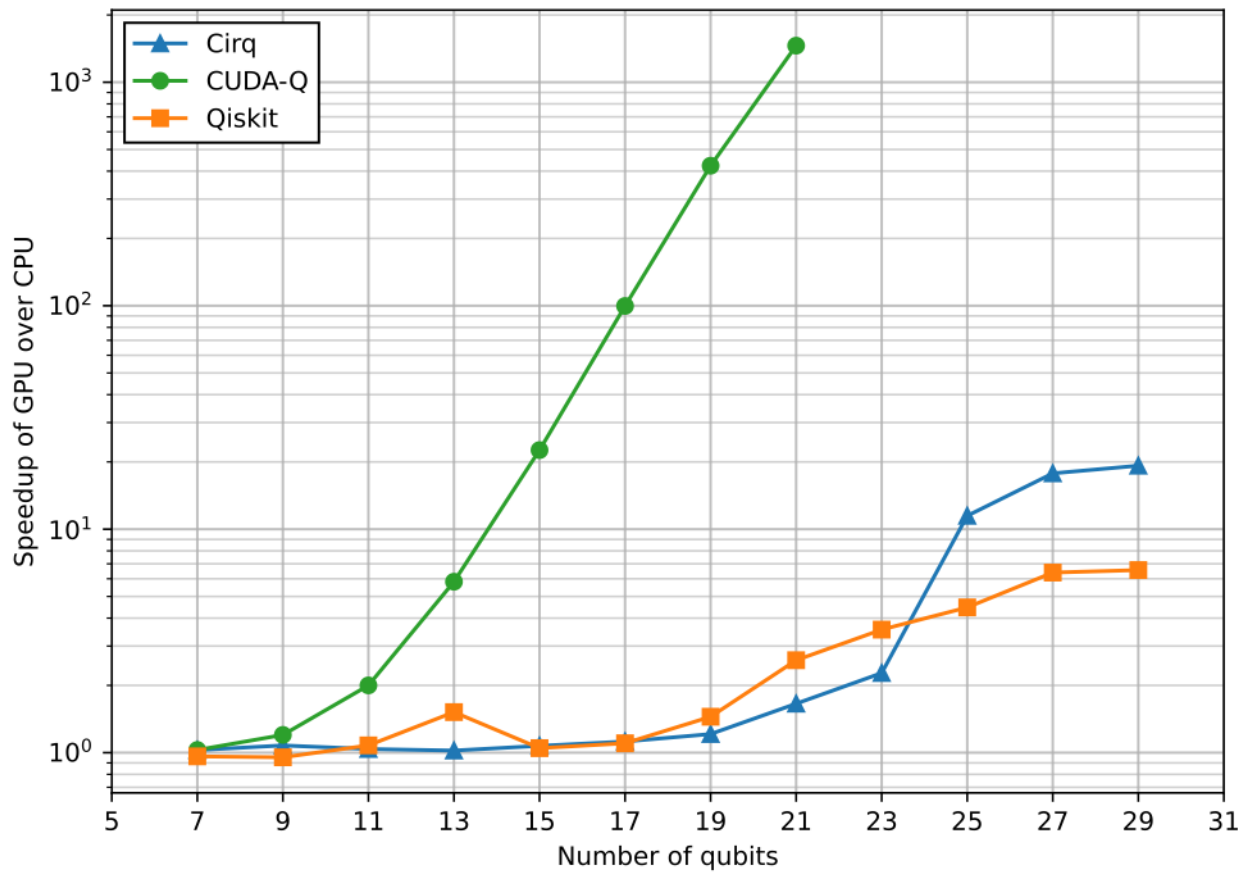

Fig. 6. GPU speedup of each framework, defined as the ratio of its CPU to its GPU average kernel runtime at equal qubit count (values > 1 favor the GPU). Ratios reflect the combined effect of GPU strength and CPU weakness; CUDA-Q's curve ends at 21 qubits because its CPU backend (qpp) was discontinued at larger sizes, and its large ratios stem largely from a weak CPU baseline.

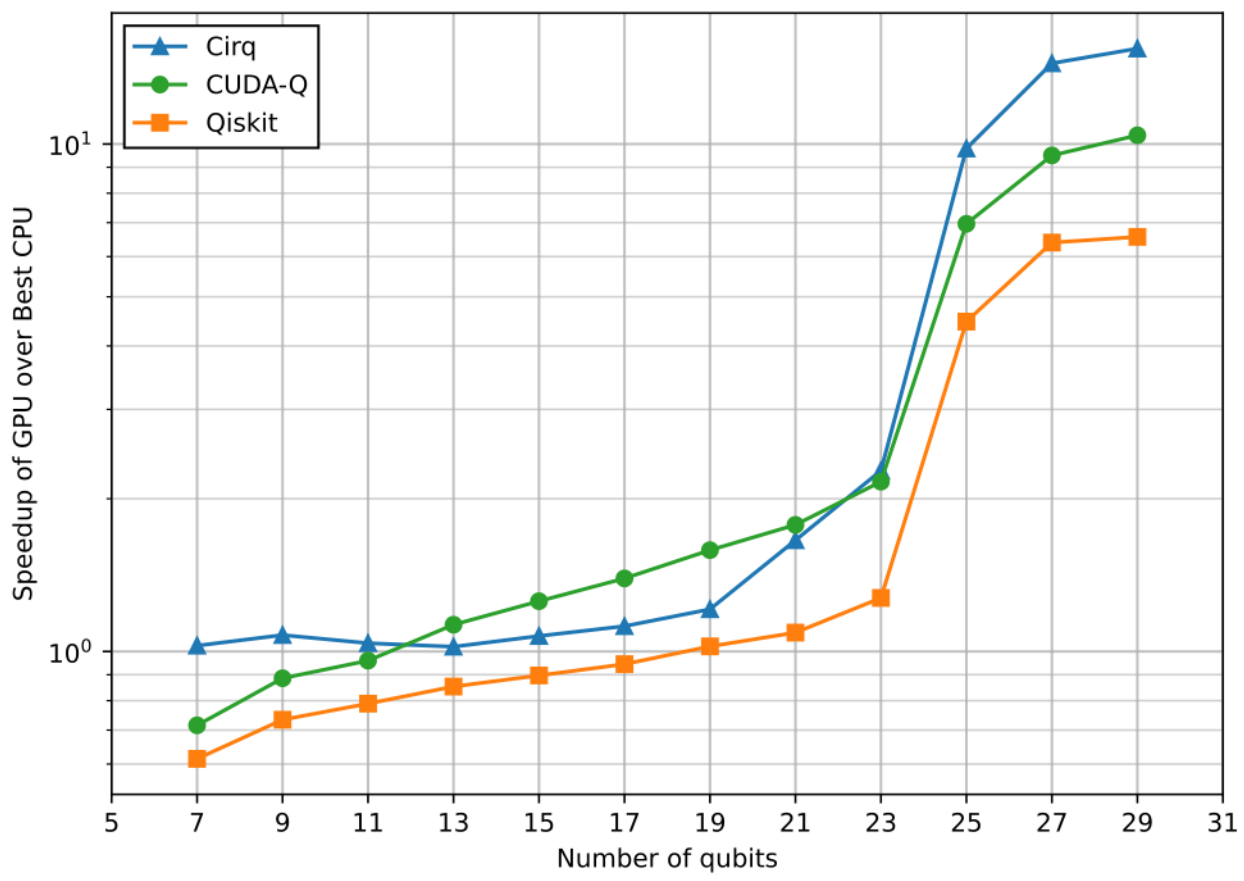

Fig. 7. GPU speedup over the best CPU backend, defined as the ratio of the fastest CPU average kernel runtime – Cirq for 7–23 qubits, Qiskit for 25–29 qubits – to each framework's GPU runtime at equal qubit count. Normalizing against the strongest CPU baseline removes the weak-baseline distortion of Fig. 6 and enables a fair cross-framework comparison.

CPU at 13 qubits for CUDA-Q, 17 qubits for Cirq, and 21 qubits for Qiskit. At the largest scale the fair metric ranks the frameworks in the same order as their absolute GPU runtimes (Fig. 5) – Cirq highest at 15× (29 qubits), then CUDA-Q at 10× and Qiskit at 6.6×, rather than in the artifact-driven order of Fig. 6. The sharpest gain for all three frameworks occurs between 23 and 25 qubits, driven by Cirq's steep, framework-specific CPU slowdown rather than by any change in GPU behavior.

No single framework dominates across all configurations. Cirq offers the strongest CPU performance at small-to-medium qubit counts and the fastest GPU runtime at large qubit counts, making it well suited for large-scale GPU simulations. CUDA-Q provides the fastest GPU execution in the medium-scale regime, but its CPU backend is impractical for large problems. Qiskit delivers the most efficient CPU multi-threading, overtaking Cirq at high thread counts and at large qubit counts, yet its GPU backend trails the other two across the full range tested. The optimal choice for execution speed therefore depends on the target configuration. In practice, framework selection also rests on criteria beyond runtime – feature coverage, supported hardware backends, programming models, and ecosystem maturity – which lie outside the scope of this study.

### B. *Future Work*

Several directions remain for future investigation. First, extending the benchmark to other classical-physics applications would test whether the performance characteristics observed here generalize beyond the acoustic wave equation. Second, multi-GPU and multi-node benchmarks would evaluate each framework's distributed-simulation capabilities; this is particularly informative because the frameworks adopt different parallelization strategies. Notably, both Cirq (via qsim) and CUDA-Q rely on NVIDIA's cuQuantum library for multi-GPU statevector simulation, whereas Qiskit Aer offers both a native chunk-based multi-GPU approach [23] and a cuQuantum-accelerated mode, allowing the performance implications of these approaches to be compared directly. Third, tensor network-based circuit simulation backends offer a complementary approach that may enable simulation at larger qubit counts than statevector methods permit, and comparing these across frameworks would provide a more complete picture of the available simulation landscape.

## Acknowledgment

The authors used generative AI code generation tools to automate the benchmarking workflow, including repeated execution of quantum circuits across multiple configurations. The core quantum circuits interfacing with each framework were developed by the authors.